\documentclass[11pt]{article}
\usepackage{moriond,epsfig}

\begin{document}
\title{Dynamics of the Inductive Single-Electron Transistor}

\author{\underline{Mika A. Sillanp\"a\"a}$^1$,
Leif Roschier$^1$, and Pertti J. Hakonen$^{1}$}

\address{$^1$Low Temperature Laboratory, Helsinki University of
Technology, Otakaari 3 A, Espoo P.O.Box 2200 FIN-02015 HUT
Finland}

\maketitle\abstracts{ Using a classical equation of motion,
dynamics of the phase is analyzed in the Inductive Single-Electron
Transistor (L-SET) which is a promising new system suitable for
quantum measurement with ultimate sensitivity and low back-action.
In a regime of nonlinear dynamics, a shift of the oscillator
resonant frequency is discovered which has a direct analogy to the
switching of a dc-biased Josephson junction into voltage state.
Results are reviewed for the predicted charge sensitivity, and it
is shown that a performance challenging the best rf-SETs is
foreseeable with the new device.}

\section{Introduction}

The Radio-Frequency Single-Electron Transistor \cite{rfset}
(rf-SET) has proven a suitable workhorse for high-frequency charge
measurements. Basically, no other device has been able to track
dynamic single-charge transport at MHz frequencies, a property
relevant especially from the point of view of characterization and
readout of superconducting qubits \cite{makhlin}. However, there
are reasons to search for an alternative technology to overcome
limitations principally due to the dissipative nature of the
rf-SET \cite{zorinrf}; (1) it is not a truly quantum-limited
detector, (2) relatively high power dissipation required for
operation heat up the sample and surrounding qubits considerably,
(3) it would be beneficial to be able to combine qubit and
detector into a single device.

Purpose of the present paper is to present an intuitive picture of
the operation a new kind of radio-frequency charge detector, the
Inductive Single-Electron Transistor (L-SET), introduced recently
by Sillanp\"a\"a \textit{et al.} \cite{lset} both theoretically
and experimentally. The L-SET is based on reactive readout of the
Josephson inductance of a superconducting SET (SSET) in a
resonator configuration. The schematics is ultimately a
non-dissipative, high-bandwidth, and sensitive electrometer with
the important property that it lacks the shot noise and thus
excessive back-action inherent in dissipative detectors due to
sequential tunneling.

\section{The L-SET circuit}

\begin{figure}
\centering \epsfig{figure=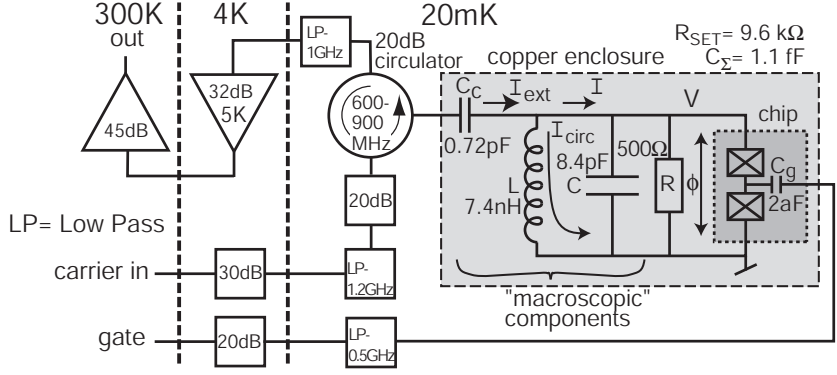, width=11cm}
\caption{Schematics of the L-SET electronics. The component values
refer to sample 2. The resonance is read by measuring amplitude
and phase of the reflection coefficient $\Gamma =
(Z-Z_0)/(Z+Z_0)$, where $Z_0 = 50 \Omega$, and $Z(L_J^{*})$ the
resonator impedance (including $C_c$).} \label{fig:circuit}
\end{figure}

We use the circuit shown in Fig. \ref{fig:circuit} where the SSET
is coupled in parallel to an $LC$ oscillator resonant at the
frequency $f_0 = 1/(2 \pi) (L C)^{-1/2}$. The resonance of the
whole system, $f_p = 1/(2 \pi) (L \parallel L_{J} C)^{-1/2} > f_0$
depends on the SSET effective Josephson inductance $(L_J^{*})^{-1}
= (2 \pi / \Phi_0)^2 E_J^{*}$ in parallel with the external
inductor $L$. Here, $\Phi_0 = h/(2 e)$ is the flux quantum. The
effective values of Josephson coupling $E_J^{*} =
\partial ^2 E(q_g, \phi) / \partial \phi^2$ and critical current
$I_0^{*} \simeq 2 e E_J^{*} / \hbar$ (not exact due non-sinusoidal
current-phase relation) are due to the quantum band structure
$E(q_g, \phi)$, where $\phi$ is the phase difference $\phi = 2 e /
\hbar \int_{0}^{t} V(t) \mathrm{d} t$ across the SSET. Since
$L_J^{*}(q_g)$ can have a substantial dependence on the gate
charge $q_g = C_g V_g$, the value of the resonant frequency $f_p$
can be used for building a sensitive detector. Its bandwidth
$\Delta f \simeq f_0 / Q_e$, where $Q_e$ is the coupled quality
factor, is typically in the range of tens of MHz.

\section{Phase dynamics}

Let us consider the SSET as a single Josephson junction which has
the current-phase relation $I_J \simeq I_0^{*} \sin(\phi)$. Using
the RCSJ-model, a classical equation of motion can be derived for
the phase difference $\phi$ across the resonator. The capacitance
is due to the shunting $C$ (Fig.\ \ref{fig:circuit}) and by the
resistor $R$ we model the small residual dissipation. The
resulting dynamics has an intuitive mechanical analog.

Current $I$ in Fig.\ \ref{fig:circuit} is divided in the junction,
$R$ and $C$,

$$I=I_0^{*} \sin(\phi) + V/R + C \dot{V} =
I_0^{*} \sin(\phi) + \hbar / (2e R) \dot{\phi} + C \hbar / (2e)
\ddot{\phi}.$$

\noindent The system is driven trough the coupling capacitor
$C_c$. The originating current $I_{ext}$ is divided, $I_{ext} = I
+ I_{circ}$, where $I_{circ}$ is the current circulating in the
loop through the external inductor $L$. It is simply related to
$\phi$ via the magnetic flux in $L$, $\Phi = \Phi_0 /2 \pi \phi =
L I_{circ}$.

Combining, we have classical non-linear equation of motion for
internal dynamics of the phase $\phi$

\begin{equation}\label{eq:eqmotion}
    C \frac{\hbar}{2e} \ddot{\phi} +\frac{\hbar}{2e R} \dot{\phi} +
    \frac{\Phi_0}{2\pi L} \phi +I_0^{*} \sin(\phi) = I_{ext},
\end{equation}

\noindent which is analogous to a mechanical particle moving in a
\emph{sinusoidally modulated parabolic potential} $V(\phi) =
\Phi_0^2/(8 \pi^2 L) \phi^2 - E_J^{*} \cos(\phi)$ and subject to a
force $I_{ext}$. Let us take a drive $I_{ext} = I_e \cos(\omega
t)$, and assume that the response is dominated by components at
the drive angular frequency, $\phi = x \sin(\omega t) + y
\cos(\omega t)$. By keeping terms of the zeroth and the first
harmonic, non-analytic equations are derived for the quadrature
amplitudes $x$ and $y$,


\begin{eqnarray}
& & - M \omega^2 x - c \omega y + k x + 2 I_0^{*} \left[ J_1(x) J_0(y) + J_1(x) J_2(y) \right] = 0 \\
& & - M \omega^2 y + c \omega x + k y - I_e + 2 I_0^{*} \left[
J_0(x) J_1(y) + J_1(y) J_2(x) \right] =  \label{eqmotsol} 0
\end{eqnarray}

\noindent where $J_n$ are Bessel functions of the first kind and
order $n$, $M = C \hbar /(2 e)$, $c = \hbar / (2 e R)$, and $k =
\Phi_0 / (2 \pi L)$. In the absence of friction, $R \rightarrow
\infty$, the response is in phase with the drive, $x = 0$, and
Eqs.\ 2, \ref{eqmotsol} simplify to

\begin{equation}
-M \omega^2 y + k y - I_e + 2 I_0^{*} J_1(y) = 0
\label{eq:nofriction}
\end{equation}

\noindent which allows for a simple graphical interpretation.

\begin{figure}
\centering \epsfig{figure=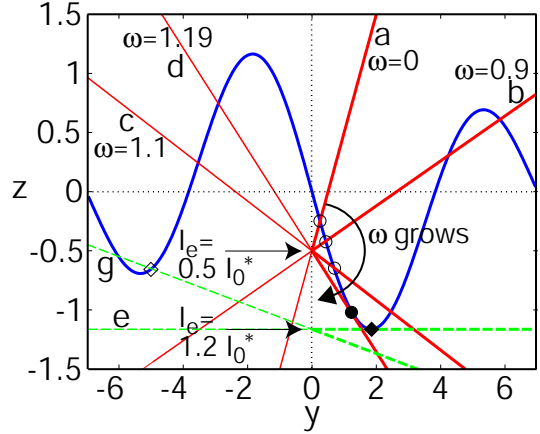, width=7cm}
\caption{Eq.\ \ref{eq:nofriction} illustrated, with $M = k =
I_0^{*} = 1$, and $\omega_0 = \sqrt{k / M}= 1$. The
smallest-amplitude solution is marked by hollow symbols, and the
resonance condition by solid symbols. For lines $e$ and $g$,
$\omega = 1.0$ and $1.05$, respectively.} \label{fig:bessel}
\end{figure}

The solution for the amplitude $y$ is the intersection of the line
$z = (k - M \omega^2) y - I_e$, and the Bessel term $z = - 2
I_0^{*} J_1(y)$. Slope of the line, $k - M \omega^2$, decreases
with increasing drive frequency $\omega$. At the resonance
frequency $\omega_R(I_e)$, the phase of the response shifts by $2
\pi$, which in this picture corresponds to \emph{change of sign}
of the solution $y$. While increasing the drive \emph{frequency}
(see Fig.\ \ref{fig:bessel}), the solid lines $a ... d$, plotted
for increasing frequency but constant drive $I_e = 0.5 I_0^{*}$,
rotate clockwise about the point $z = - I_e$. At relatively low
drive, the favored solution which has the smallest absolute value
and hence the lowest energy, changes sign at the frequency
$\omega_R \simeq 1.19 \omega_0$ where the line $d$ tangents the
Bessel function at the point marked with solid circle. We note
that the line $d$ has a clearly negative slope, which corresponds
to a relatively large value of $\omega_R$.

While increasing the drive, at the particular value $I_e \simeq
1.2 I_0^{*}$ the sign of the smallest solution changes when the
line is \emph{horizontal}, as depicted by the line $e$, which
corresponds to \emph{a smaller} frequency than in the previous
case. At this $\omega$, the solution changes from $y \simeq 1.8$
(solid diamond) to $y = -\infty$, and approaches again zero (line
$g$) at higher frequency. The infinity is due to absence of
dissipation. At still higher drive, the resonance frequency does
not change in this simplified picture. Note also that if the
Bessel term due to the Josephson contribution $\rightarrow 0$, the
resonance condition corresponds to a horizontal line as well.

We re-formulate the statements: At the "critical drive" $I_e
\simeq 1.2 I_0^{*}$ the resonance frequency $\omega_R$ shifts from
a high $\omega_R = 2 \pi f_p$ which includes the
Josephson-contribution, to the lower $\omega_R = 2 \pi f_0$ which
is determined only by $L$ and $C$.

This shift of resonance frequency of the L-SET oscillator at a
certain "critical power" $P_c$ of the microwave drive is in a
direct analogy to the switching of a DC-biased Josephson junction
into voltage state \cite{yale}. Of course, a similar phenomenon
should take place equally well in a single junction than the SSET.

\section{Experimental results}

Results of two samples (see Table \ref{tb:samples}) are discussed.
We focus here on the phase dynamics of sample 1. Since its $E_C$
was comparable to temperature, it did not work as a good
electrometer, however, it was therefore rather close to a
classical junction supposed in the calculations. Sample 2
(discussed in Ref.\ \cite{lset}, see also section
\ref{sec:sensit}) was a sensitive detector, with a gate shift of
the resonance frequency of 15 MHz.

\begin{table}
\centering
\begin{tabular}{|l||l|l|l|l|l|l|l|l|}
\hline sample & $R_{SET}$ (k$\Omega$) & $E_J$ (K) & $E_C$ (K)
&$L_J^{*}$ (nH) & $L$ (nH) & $C$ (pF) & $C_c$ (pF) & $Q_e$ \\
\hline \hline
  1 & 4.2 & 3.5 & 0.17 & 6 & 3 & 23 & 0.72 & 13 \\ \hline
  2 & 9.6 & 1.6 & 0.92 & 16 & 7.4 & 8.4 & 0.72 & 18 \\ \hline
\end{tabular}
\caption{Parameters of the samples and their tank $LC$ oscillators
discussed in the text. $E_J = h \Delta /(8 e^2 \frac{1}{2}
R_{SET})$ is the single-junction Josephson energy, and $E_C =
e^2/(2 C_{\Sigma})$ is the charging energy.} \label{tb:samples}
\end{table}

\begin{figure}
\centering \epsfig{figure=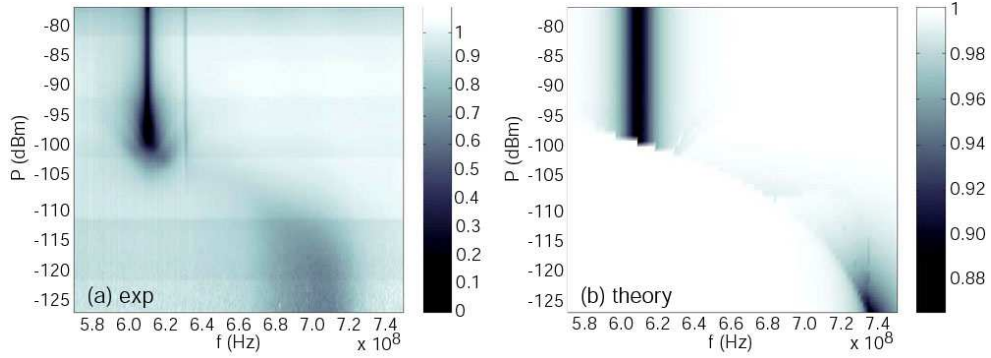, width=13cm}
\caption{Frequency response ($| \Gamma |$) of sample 1 as a
function of increasing drive amplitude. (a) experiment; (b)
numerical calculation using Eqs.\ 2, \ref{eqmotsol} and circuit
relations for $Z$. The gray scales in (a) and (b) are scaled from
1 to the smallest $| \Gamma |$ in each case.}
\label{fig:freqpower}
\end{figure}

In Fig.\ \ref{fig:freqpower} (a) we plot the frequency response
for sample 1 as a contour, a suitable representation for the
dynamics as a function of drive strength. The shift of resonance
frequency from the plasma resonance at $f_p = 710$ MHz to the tank
resonance at $f_0 = 610$ MHz is clearly visible at a critical
power $P_c \simeq - 102$ dBm. A prediction from our formalism
which uses only independently determined parameter values, Fig.\
\ref{fig:freqpower} (b), agrees qualitatively with the features of
the experimental data (yet the resonance depth is different due to
probably too small value of $C_c$).

\section{Charge sensitivity} \label{sec:sensit}

The bottleneck for detector sensitivity in the present schematics
is the preamplifier having a noise temperature of 5 K. For sample
2 in the discussed operation mode of small plasma oscillations, we
measured a charge sensitivity $s_q \simeq 2 \times
10^{-3}$e$/\sqrt{\mathrm{Hz}}$. At higher drive which corresponds
to several periods of the $\cos(\phi)$ Josephson term in the
potential ("anharmonic mode"), such that the transduction is not
due to tuning of $f_p$ but rather, due to tuning of non-linear
phase dynamics, a better sensitivity of $1.4 \times
10^{-4}$e$/\sqrt{\mathrm{Hz}}$ was achieved. Since the sensitivity
is predicted to improve roughly as $(E_J / E_C)^{-1}$, the record
numbers of the rf-SET, $s_q \sim 3 \times
10^{-6}$e$/\sqrt{\mathrm{Hz}}$, should be beatable at
experimentally accessible values of $E_J / E_C \sim 0.4$ in the
anharmonic mode, or with $E_J / E_C \sim 0.15$ in the plasma
oscillation mode, both using an Al SSET in the present
experimental scheme \cite{tbb}.

\section*{Acknowledgments}
Collaboration with T. Heikkil\"a, G. Johansson, R. Lindell, H.
Sepp\"a, and J. Viljas is gratefully acknowledged. This work was
supported by the Academy of Finland and by the Large Scale
Installation Program ULTI-3 of the European Union.

\end{document}